# Layered material GeSe and vertical GeSe/MoS$_2$ p-n heterojunctions

Wui Chung Yap[1]*, Zhengfeng Yang[1]*, Mehrshad Mehboudi[2]*, Jia-An Yan[3], Salvador Barraza-Lopez[2], and Wenjuan Zhu[1]#

[1]Department of Electrical and Computer Engineering, University of Illinois at Urbana-Champaign. Urbana, IL 61801, USA

[2]Department of Physics, University of Arkansas. Fayetteville, AR 72701, USA

[3]Department of Physics, Astronomy & Geosciences, Towson University. Towson, MD 21252, USA

## Abstract

Beyond graphene, transition metal dichalcogenides (TMDCs) and black phosphorus (BP), group-IV monochalcogenides are emerging as a unique class of layered materials. We report in this paper experimental and theoretical investigations of germanium selenide (GeSe) and its heterostructures. We find that GeSe has a prominent anisotropic electronic transport with maximum conductance along the armchair direction. Density functional theory (DFT) calculations reveal that the effective mass along the zigzag direction is 2.7 times larger than that in the armchair direction, and the anisotropic effective mass explains the observed anisotropic conductance. The crystallographic direction of the GeSe is confirmed by angle-resolved polarized Raman measurement, which is further supported by calculated Raman tensors for the orthorhombic structure. Novel GeSe/MoS$_2$ pn heterojunctions are created, taking advantage of the natural p-type doping in GeSe and n-type doping in MoS$_2$. The temperature-dependence of the junction current measurement reveals that GeSe and MoS$_2$ form a type II band alignment with a conduction band offset of ~ 0.234 eV. The anisotropic conductance in GeSe may enable a new series of electronic and optoelectronic devices such as plasmonic devices with resonance frequency continuously tunable with light polarization direction and high-efficiency thermoelectric devices. The unique GeSe/MoS$_2$ pn junctions with type II alignment will be an essential building block for vertical tunneling field-effect transistors for low power applications. This new p-type layered material GeSe can also be combined with n-type TMDCs to form heterogeneous complementary metal oxide semiconductor (CMOS) circuits.

------------------------------------------------------------

* These authors contributed equally to this paper.

# Corresponding author's email: wjzhu@illinois.edu

Layered group-IV monochalcogenides are a newly emerging materials platform. As compared to graphene, transition metal dichalcogenides (TMDCs) and black phosphorus (BP), layered monochalcogenides such as SnS, SnSe, GeS and GeSe have many unique electrical, thermal, and optical properties that could prove useful for diverse applications. In particular, layered monochalcogenides have an orthorhombic (distorted rock-salt) structure and feature anomalously high Grüneisen parameters, which lead to ultralow thermal conductivity and an exceptionally high thermoelectric figure of merit,[1, 2] which makes them promising for thermoelectric applications. At high temperatures, layered monochalcogenides undergo a displacive phase transition from a lower symmetry Pnma space group to a higher symmetry Cmcm space group,[1, 3] which makes them an interesting candidate for phase change memory. The bandgap of layered group-IV monochalcogenides is in the range of 0.5 to 1.5eV,[4] lining up fairly well with the solar spectrum, which makes them attractive for solar cells and photodetectors too. In addition, the effective masses of some layered monochalcogenides are much smaller than those of TMDCs,[5] which can lead to higher carrier mobilities for electronic applications. In addition, unlike phosphorene, the valleys for monochalcogenides are away from the Gamma-point, and carriers at each valley can be separately excited with linearly-polarized light.

Among layered monochalcogenides, GeSe is a narrow bandgap semiconductor, which is particularly attractive for near-infrared (NIR) photodetectors and electronic tunneling devices. The crystal structure of GeSe and the coordinates used in this paper are illustrated in Figure 1. In the bulk, GeSe has an orthorhombic structure with Bernal stacking containing eight atoms per unit cell. The lattice structure can be viewed as a distorted rocksalt structure forming double-layer planes perpendicular to the $a_3$-axis.[6] Each atom has three strongly bonded neighbors within its own puckered layer, and three more distant neighbors not forming covalent bonds in adjacent layers.[6] GeSe has an indirect bandgap of 1.08 eV in bulk [7] and a direct bandgap of ~1.7 eV in monolayers.[5, 8] Single crystal GeSe with stoichiometric composition has hole mobilities of 95 $cm^2$/V-s at 300K and 663 $cm^2$/V-s at 112K.[9] Theoretically, it has been predicted that the average hole mobility for monolayer GeSe is as high as $1.1 \times 10^3 cm^2$/V-s at 300K.[10] It has been reported that GeSe has a high photoresponsivity along the $a_3$ (perpendicular to the plane) direction.[11, 12] Despite the intense research on bulk GeSe crystals and theoretical calculations on monolayers[5-13], experimental investigation of the anisotropic current transport of GeSe and GeSe heterostructures is still unrealized.

Motivated by the novel features and functional possibilities in monochalcogenides, here we discover an anisotropic current transport in GeSe and demonstrate the properties of novel GeSe/MoS$_2$ heterojunctions. We find that GeSe has maximum conductance along the armchair direction. Basic electronic structure calculations reveal that the effective mass is orientation-dependent at the valence valley edge: it is largest along the $a_2$ (zigzag) direction, and smallest along the $a_1$ (armchair) direction. The smaller mass along the armchair direction results in the highest mobility and conductivity experimentally observed along that direction. The crystal orientation of GeSe was additionally confirmed by angle-dependent polarized Raman measurements, which yield an orientation of the in-plane lattice parameters consistent with the predictions based from effective-mass arguments. Here the GeSe flakes were obtained by exfoliation with thickness ranges from 14 nm to 230 nm measured by atomic force microscopy (AFM). Taking advantage of the p-type doping in GeSe, we then fabricated pn junctions based on MoS$_2$/GeSe heterostructures. The temperature dependence of the junction current is measured, from which the band offset in MoS$_2$/GeSe was determined. The results reveal a type II band alignment between MoS$_2$ and GeSe. These naturally formed pn junctions with type II band alignment are very promising for tunneling field effect transistor (TFET) applications.

**Results:**

**1. Anisotropic electronic transport**

The anisotropic current transport in GeSe was studied by angle-resolved DC conductance measurements. The optical image of a GeSe device is shown in Figure 2a. Twelve electrodes with angular spacing of 30° between electrodes were fabricated on the GeSe flake. We performed DC conductance measurements by applying electric fields across pairs of electrodes positioned at angles of 180° apart. Here we select the direction parallel to the electrode pair labelled T5 as the 0° reference direction and define $\theta_c$ as the angle between the measured electrode and the 0° reference direction. Figure 2b shows a polar plot of the conductance as a function of angle $\theta_c$. The maximum conductance is along the electrode pair T5 at $\theta_c = 0°$ angle and the minimum conductance is along the orthogonal electrode pair T2 at $\theta_c = 90°$ angle. As the temperature decreases from 300K to 200K, the conductance reduces dramatically. This mainly due to the reduced carrier density generated thermally at lower temperatures. Mobility degradation due to increased Columbic scattering at lower temperatures is another possible factor that can result in a reduction of the channel conductance. In all cases, conductance reaches a maximum at 0° angle regardless of temperature.

Calculations reveal that the anisotropy on the conductivity is a direct signature of the anisotropy of the hole valley. The band structure shown in Figure 2c (see Methods) illustrates the bands responsible for charge transport in a p-doped sample.

Two valleys whose energy maxima are located at $k_0 = \left(\pm 0.3552 \frac{2\pi}{a_1}, 0, 0\right)$ along the Γ-X high symmetry line contribute the hole current within a 0.1 eV range (the hole valley maxima at $k_0 = \left(+0.3552 \frac{2\pi}{a_1}, 0, 0\right)$ is explicitly indicated with a red vertical arrow in the band structure plot). The blue horizontal dashed line in Figure 2c indicates the energy location of the valley maxima, which corresponds to the Fermi energy for a lightly p-doped sample. The inset in the band structure plot, Figure 2c, displays the high-symmetry points in the Brillouin zone, and two yellow circumferences with radius $0.07\pi/a_1$ centered around these two valley maxima within the first Brillouin zone.

In Figure 2d, the focus is placed on the electronic structure at the valley. The shape of the hole valley $E_{hv}(k_x, k_y)$ was calculated on a polar k-point mesh (r, θ) centered at the hole valley energy maxima $k_0 = \left(\pm 0.3552 \frac{2\pi}{a_1}, 0, 0\right)$, with $a_1 = 4.557$Å the longest in-plane lattice constant, with a radius $r = 0.07 \frac{\pi}{a_1}$, a radial resolution of $r/25$, and angular resolution $\Delta\theta = 5°$ that is indicated by a gray line in the inset (see Ref. [14] for structural details). The energy as a function of $|k - k_0|$ cut at three angles is plotted in Figure 2d, and the angular anisotropy of the hole valley is clearly revealed.

Now, the explicit expression for the mobility µ has an isotropic dependence on the deformation potential, the lattice parameters $a_1$ and $a_2$, and temperature which can be encompassed on a single parameter $c$.[15-17] The only angular dependence on µ arises from the product of effective masses:

$$\mu(\theta) = \frac{c}{m^*(\theta)\sqrt{m^*(\theta)m^*(\theta+90)}} \quad (1)$$

which must be computed. Here $\theta$ is defined as the angle between the current transport direction and the armchair direction in the GeSe crystal, and $c$ is an isotropic parameter. The angle-dependent hole mass is estimated from plots like the one shown in Figure 2d from:

$$\frac{1}{m^*(\theta_i)} \equiv -\frac{1}{\hbar^2}\frac{\partial^2 E_{hv}(k-k_0)}{\partial k(\theta_i)^2} \quad (2)$$

where $k(\theta_i)$ implies that the derivative is being taken along the ray $\theta_i \equiv i\Delta\theta$ centered at $k_0$ (parabolic fits to the band dispersion are shown in dashed lines in Figure 2d). Figure 2e shows the effective mass as a function of the angle $\theta$ in the polar plot. We can see that the effective mass is lightest when $\theta = 0°$, i.e along the armchair direction, and is heaviest when $\theta = 90°$, i.e along the zigzag direction.

The hole conductivity is proportional to the mass anisotropy through:

$$\sigma(\theta) = \int_{E_{Max}}^{E} 4\rho(E)|e|\mu(\theta)dE \quad (3)$$

where $\rho(E)$ is the charge density at energy $E$ below the band edge, and the factor of four accounts for valley and spin degeneracy. The point is that Equation 1 provides the only angular anisotropy to the conductivity. In Figure 2b we superimpose the experimental data to the continuous curves obtained from Equation 1 using a single, temperature-dependent, scaling factor $c$ that encompasses spin and valley degeneracy, charge carrier density and deformation potential, i.e., variables which only provide isotropic contributions to the conductivity. The temperature dependence of the parameter c is plotted in Figure 2f. We can see that ln(c) versus reciprocal of temperature roughly follows a straight line, which is consistent to the fact that the density of the intrinsic carriers generated thermally has exponential dependence of the reciprocal of the temperature: $n_i \propto exp(-E_g/2k_BT)$, where $E_g$ is the band gap of the GeSe, and $k_B$ is Boltzmann constant.

This way, the experimental data thus reveals the valley mass anisotropy directly, so it can be used to determine the two orthogonal in-plane crystalline directions. Conductance maxima aligns with $\theta = 0°$ which corresponds with the armchair direction, and minima aligns with $\theta = 90°$, corresponding to the zigzag direction.

The anisotropic conductance in GeSe may enable a new series of electronic and optoelectronic devices, such as plasmonic devices tunable by light polarization, and high-efficiency thermoelectric devices. It has been found that the plasmon resonance frequency of nano-disks built on anisotropic materials such as BP, is dependent on the incident light polarization direction[18-20]. Plasmon devices based on anisotropic materials will have additional knobs to dynamically adjust resonance frequency during operation as compared to isotropic materials such as TMDCs and graphene. In addition, the maximum heat and electron transport directions in anisotropic materials can be orthogonal, which gives a high ratio of electrical to thermal conductivity and a high thermoelectric figure-of-merit.[21]

## 2. Polarized Raman and crystal orientation

To further determine the crystal directions of the GeSe flakes, we performed angle dependent polarized Raman measurements on our devices. Figure 3a shows the optical setup employed. The incident light polarizer is located behind the laser, while the receiver polarizer is placed in front of the spectra analyzer. The polarizations of the two polarizers are both parallel to the horizontal direction of the sample stage. A half-wave plate is placed between the incident light polarizer and the sample. By rotating the wave plate, the angle between the polarization of the incident light on the sample and the reference direction, α, can be tuned from 0° to 360°. The definition of the angles is illustrated in Figure 3b. The angle between the receiver polarizer and the reference direction (electrode T5) is defined as β. The receiver polarizer direction, i.e., the angle β, is unchanged during measurements. In this experiment, β=17°. The Raman measurements shown next follow the original angular disposition of the sample, and the values to be reported should be subtracted by 17° in order to compare with conductivity data discussed in previous Section. A typical Raman spectrum is shown in Figure 3c. The clearly observed peaks at 83 and 188cm$^{-1}$ are associated with the $A_g^1$ and $A_g^2$ modes, while the peak at 151cm$^{-1}$ is assigned to $B_{3g}$ mode; a result in good agreement with previous work.[22, 23] Figure 3d shows the measured and fitted intensity of $A_g$ and $B_{3g}$ modes as a function of the angle between the incident light polarization and the receiver polarizer, α. We can see that the intensities of $A_g$ and $B_{3g}$ modes have a variation period of 180°.

To better understand the experimental results, the Raman scattering light intensity, I, is modeled. In our system, the X axis is defined as the cross-plane direction, the Y axis is along the armchair direction, and the Z axis is along the zigzag direction. The incident light angle, $\theta_{in}$, is defined as the angle between the incident light polarization and the armchair direction. We can see that the incident light polarization angle is $\theta_{in} = \alpha + \beta + \gamma$, and the scattered light polarization angle is $\theta_s = \beta + \gamma$. In this system, the incident light polarization can be expressed as $e_i = (0, \cos\theta_{in}, \sin\theta_{in})$, while the scattered light polarization $e_s = (0, \cos\theta_s, \sin\theta_s)$. GeSe in its orthorhombic phase has a space group of $D_{2h}^{16}$, and the Raman tensor of Raman modes $A_g$, $B_{3g}$ are given as follows:[24, 25]

$$R(A_g) = \begin{pmatrix} A & 0 & 0 \\ 0 & B & 0 \\ 0 & 0 & C \end{pmatrix} \qquad R(B_{3g}) = \begin{pmatrix} 0 & 0 & 0 \\ 0 & 0 & F \\ 0 & F & 0 \end{pmatrix} \qquad (4)$$

Then, the Raman scatter intensity, I, can be expressed as:
$$I(B_{3g}) = F^2 \sin^2(\theta_{in} + \theta_s) \qquad (5)$$
$$I(A_g) = (B \cos\theta_{in} \cos\theta_s + C \sin\theta_{in} \sin\theta_s)^2 \qquad (6)$$

By fitting $I(B_{3g})$ and $I(A_g)$, we get two solutions. If $|B| < |C|$, the angle between the receiver polarizer and the armchair direction, $\gamma = 15°$, i.e., the maximum conductance is roughly along the armchair direction determined by Raman. If $|B| > |C|$, $\gamma = 105°$, i.e., the maximum conductance is along the zigzag direction. The symbols in Figure 3d are experimental results and the lines are fittings using Equation 5 and 6.

The two calculated Raman-active $A_g^1$ and $A_g^2$ modes correspond to experimental observations at 72 cm$^{-1}$ and 186 cm$^{-1}$, respectively. Their normalized Raman tensors are as follows:

$$R(A_g^1) = \begin{pmatrix} 1.0 & 0 & 0 \\ 0 & -0.707 & 0 \\ 0 & 0 & -0.877 \end{pmatrix} \quad (7)$$

$$R(A_g^2) = \begin{pmatrix} 1.0 & 0 & 0 \\ 0 & 1.212 & 0 \\ 0 & 0 & 1.691 \end{pmatrix} \quad (8)$$

In both cases, the elements in the tensor satisfy $|B| < |C|$. Therefore, we conclude that the angle between armchair direction and the conductance reference direction is $\gamma = 15°$, as illustrated in Figure 3e. This indicates that the armchair direction of the GeSe crystal determined by Raman is roughly along the maximum conductance direction. This small offset angle is due to the resolution of the angle-resolved conductance which is limited to 30° in this experiment. This result is consistent with theoretical calculation on effective mass and mobility. Thus, Raman analysis provides additional validation for the actual orientation of the experimental sample.

## 3. GeSe transistors

Figures 4a and 4b show the $I_D V_G$ characteristics of MoS$_2$ and GeSe field effect transistors (FETs), respectively. The insets show the log scale of the $I_D V_G$ characteristics. The MoS$_2$ $I_D V_G$ plot shows typical n-channel MOFET characteristics, while the GeSe $I_D V_G$ plot resembles that of a p-channel MOFET. These devices do not have any intentionally doping. This natural n-type doping in MoS$_2$ and p-type doping in GeSe may attribute to the low lying conduction band in MoS$_2$ and high lying valence band in GeSe. [26] The field effect mobility for MoS$_2$ is extracted as ~30 cm$^2$/Vs, consistent with mobility values of multilayer MoS$_2$ FETs reported previously, [27-35] while the extracted field effect mobility for GeSe is ~1 cm$^2$/Vs. This extraction did not take into account the contact resistance, and the true carrier mobility will be higher. Most of the 2D materials are naturally n-type, such as MoS$_2$, WS$_2$, MoSe$_2$, SnS$_2$.[28, 36-39] To form pn junctions or to make complementary transistors requires p-type doping as well. Currently the doping techniques in 2D and layered materials are still very immature. GeSe is naturally p-type doped and thermodynamically stable in air, providing an essential building block for n-channel (p-type substrate) transistors, complementary circuits, and tunneling field effect transistors.

## 4. GeSe/ MoS$_2$ heterostructures

Using GeSe flakes, we fabricated MoS$_2$/GeSe heterojunctions. Figure 5a shows an optical image of the device. Figure 5b shows the junction current measured at various temperatures from 296K to 220K. The room temperature IV curve resembles a typical p-n diode characteristic, with an exponential increase of current at forward bias (>100 nA) and a small current at reverse bias (~1 nA), showing that GeSe/MoS$_2$ has rectifying characteristics. The temperature dependence of the junction current can be understood using the thermionic emission model,[40]

$$I_{DS} = I_{TE} \exp\left(\frac{qV}{nkT}\right)\left[1 - \exp\left(-\frac{qV}{kT}\right)\right] \quad (7)$$

$$I_{TE} \propto T^2\left[\exp\left(-\frac{q\phi_B}{kT}\right)\right] \quad (8)$$

where $I_{TE}$ is the saturation current, $k$ is the Boltzmann constant, $V$ is the applied voltage, $T$ is the temperature, and $q$ is the electron charge. The ideality factor $n$ and the band-offset $\phi_B$ at the heterojunction can be extracted. From the semi-log plot of $I_{DS}/[1 - e^{-eV/kT}]$ versus $V$ (Figure 5c), the slopes of the linear regions give the ideality factor, $n$, values between 2.37 and 2.64 for all temperature values, and the intercepts give $\ln(I_{TE})$ for each temperature. Using values of $I_{TE}$ for different temperatures, a semi-log plot of $\frac{I_{TE}}{T^2}$ versus $\frac{q}{kT}$ (Figure 5d) gives a band-offset value of ~0.234 eV. This indicates that the GeSe/MoS$_2$ heterostructure forms a pn junction with type II band alignment, which is essential for vertical TFET devices.

**Conclusion:**
In summary, we have introduced GeSe as a new anisotropic material with a natural p-type doping and GeSe/MoS$_2$ as a novel pn heterostructure to the layered materials family. More specifically, we found that GeSe has a maximum conductance along the armchair direction based on polarized Raman and angle-resolved conductance measurements. DFT calculations reveal that the effective mass is lightest along the armchair direction and heaviest along the zigzag direction, which is consistent with the experimental observation. A novel p-n heterojunction consisting of GeSe and MoS$_2$ was fabricated and systematically characterized at various temperatures. Our results indicate a type II band alignment between GeSe and MoS$_2$ is formed with conductance band offset of 0.234 eV.

The anisotropic properties of GeSe allow a new degree of freedom for designing novel photonic and electronic devices, such as polarizers, polarization sensors, and plasmonic devices where the plasma resonant frequency is tunable by the incident light polarization direction. In addition, the high thermoelectric figure-of-merit in GeSe due to the anisotropic electronic and thermal conductivity make it a promising candidate for thermoelectric applications. Most importantly, the type II band alignment between GeSe and MoS$_2$ and complementary natural doping in these two materials can enable vertical TFETs with abrupt pn junction and short screening tunneling length. These TFETs will have the characteristic of super-steep subthreshold swing and are attractive for low power applications. Moreover, this new p-type layered material, GeSe, can be combined with n-type TMDCs and enable complementary metal oxide semiconductor circuits, such as inverters, oscillators, NAND gates, and NOR gates. These are just a few promising applications for GeSe that we have identified, while an entire family of new device concepts that can utilize the unique properties of this new intriguing material still wait for our exploration.

**Methods:**
Multilayer GeSe flakes were exfoliated from bulk crystals using adhesive tape and transferred onto a SiO$_2$/Si substrate. GeSe/MoS$_2$ heterostructures were formed by aligned dry transfer technique. [41] Multilayer MoS$_2$ flakes were first exfoliated from bulk crystals and transferred onto a viscoelastic stamp supported by a glass slide. The glass slide was then positioned using a mechanical micromanipulator stage and a microscope to align the flake to the position of the GeSe, and the $MoS_2$ flake was transferred by pressing and slowly releasing the viscoelastic stamp. The electrodes were formed by e-beam lithography, metal deposition and lift-off. The electrodes consisting of 30nm titanium and 20nm gold were deposited by electron-beam evaporation. The polarized Raman measurement were carried out using Horiba LabRAM HR 3D-capable Raman spectroscopy imaging system. The angle-resolved conductance and temperature dependence of junction current were measured using Lakeshore cryogenic-free Hall probe-station and Agilent parameter analyzer B1500.

In computing the effective mass, the Perdew-Burke-Eherenzoff exchange-correlation potential [42] was employed. Calculations that employ PAW pseudopotentials [43] were carried out with the VASP code,[43, 44] and lattice parameters turned out to be $a_1$=4.557, $a_2$=3.918, and $a_3$=11.640 Å.

DFT calculations of Raman tensors were performed using the Quantum ESPRESSO (QE) code [45] with the Perdew-Zunger (PZ) local density approximation (LDA) exchange-correlation functional. Norm-conserving pseudopotentials for Ge and Se were employed for the description of interactions between core and valence electrons. The Raman tensor for each phonon mode was then obtained using density-functional perturbation theory as implemented in QE. A similar method has been applied to study the effect of strain on the Raman spectra in monolayer $MoS_2$ [46].

**Acknowledgement:**
The authors would like to thank Joseph Lyding and Arend van der Zande of the University of Illinois at Urbana-Champaign (UIUC) for insightful discussions. The authors at UIUC would like to acknowledge the NSF support through Grant ECCS#1611279, and the Arkansas-based authors acknowledge funding from the DOE (DE-SC0016139). Effective mass calculations (M.M. and S.B.L.) were performed at SDSC's Comet (Grant XSEDE TG-PHY090002).

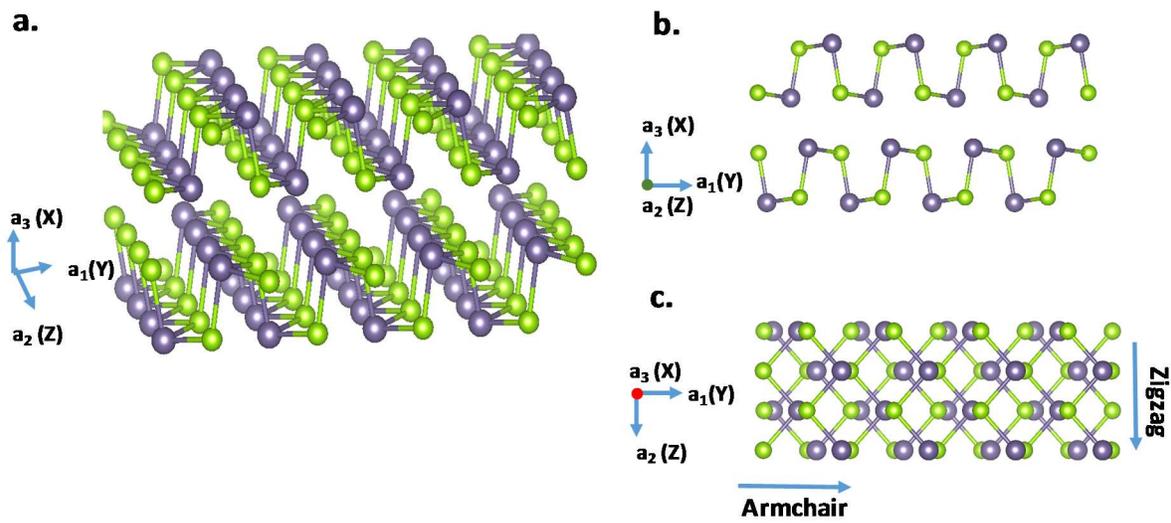

Figure 1. Crystal structure of germanium selenide, represented in (a) prospective view, (b) side view, and (c) top view. (All lattice vectors form 90 degree angles, and $a_3 > a_1 > a_2$).

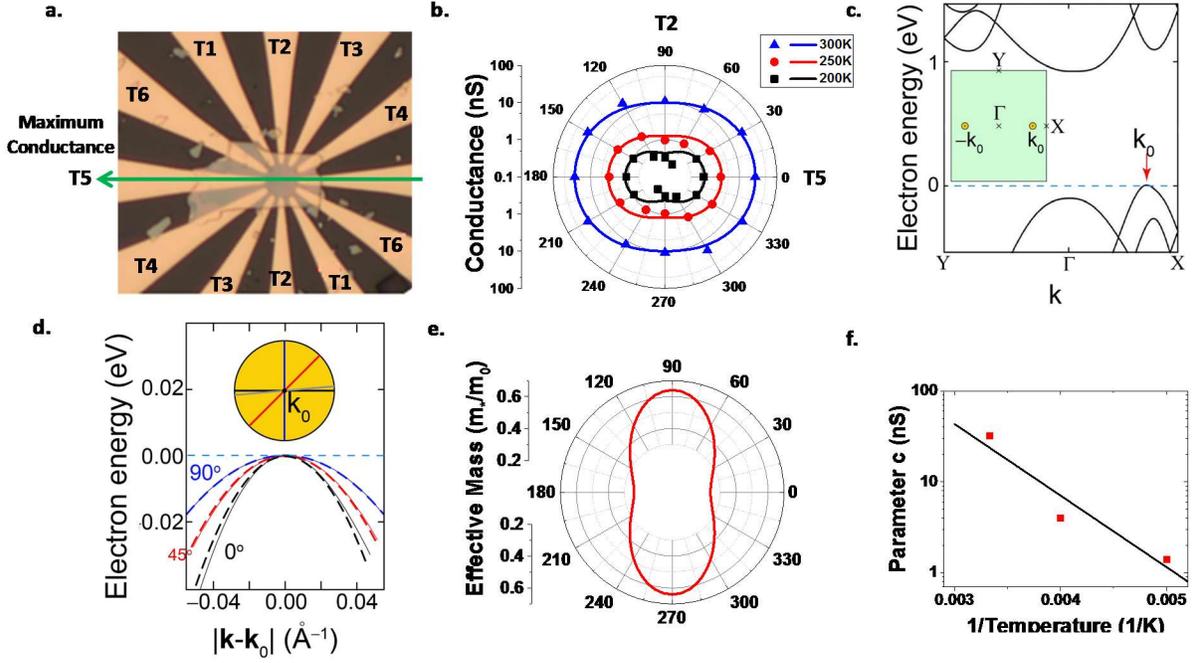

Figure 2. Anisotropic conductance in GeSe. (a) Optical image of GeSe flake with 12 electrodes spaced at 30º apart. The green line along transistor T5 is chosen as the 0º reference direction for this sample. The maximum conductance is along transistor T5. (b) DC conductance along the six directions at three temperatures (300K, 350K and 200K) and plotted in polar coordinates. The symbols are experimental results and lines are calculated results. Here the angle, $\theta_c$, is between measured electrode and reference line along T5 transistor direction. (c) Band structure of GeSe. The hole valley maxima at $k_0$ is indicated with a red vertical arrow in the plot. The blue horizontal dashed line indicates the energy location of the valley maxima, which corresponds to the Fermi energy on a lightly p-doped sample. The inset displays the high-symmetry points in the Brillouin zone, and two yellow circumferences with radius $0.07\pi/a_1$ centered around these two valley maxima within the first Brillouin zone. (d) The hole valley anisotropy is revealed by angular band structure cuts at increasing angles. Solid lines correspond to the numerical data, and the dashed lines represent quadratic fittings leading to the mass estimate $E(k, \theta) = \frac{\hbar^2 |k=k_0|}{2m^*(\theta)}$. Here θ is the angle referring to the armchair direction. The inset shows the in-plane view of the k-space around the valley maxima; the shadow gray line indicates the 5 degree angular resolution employed in calculations (e) Effective mass in units of free electron mass, $\frac{m^*}{m_0}$, as a function of angle θ. (f) Fitting parameter c as a function of reciprocal of temperature (1/T). The line is the linear fitting of ln(c) versus 1/T.

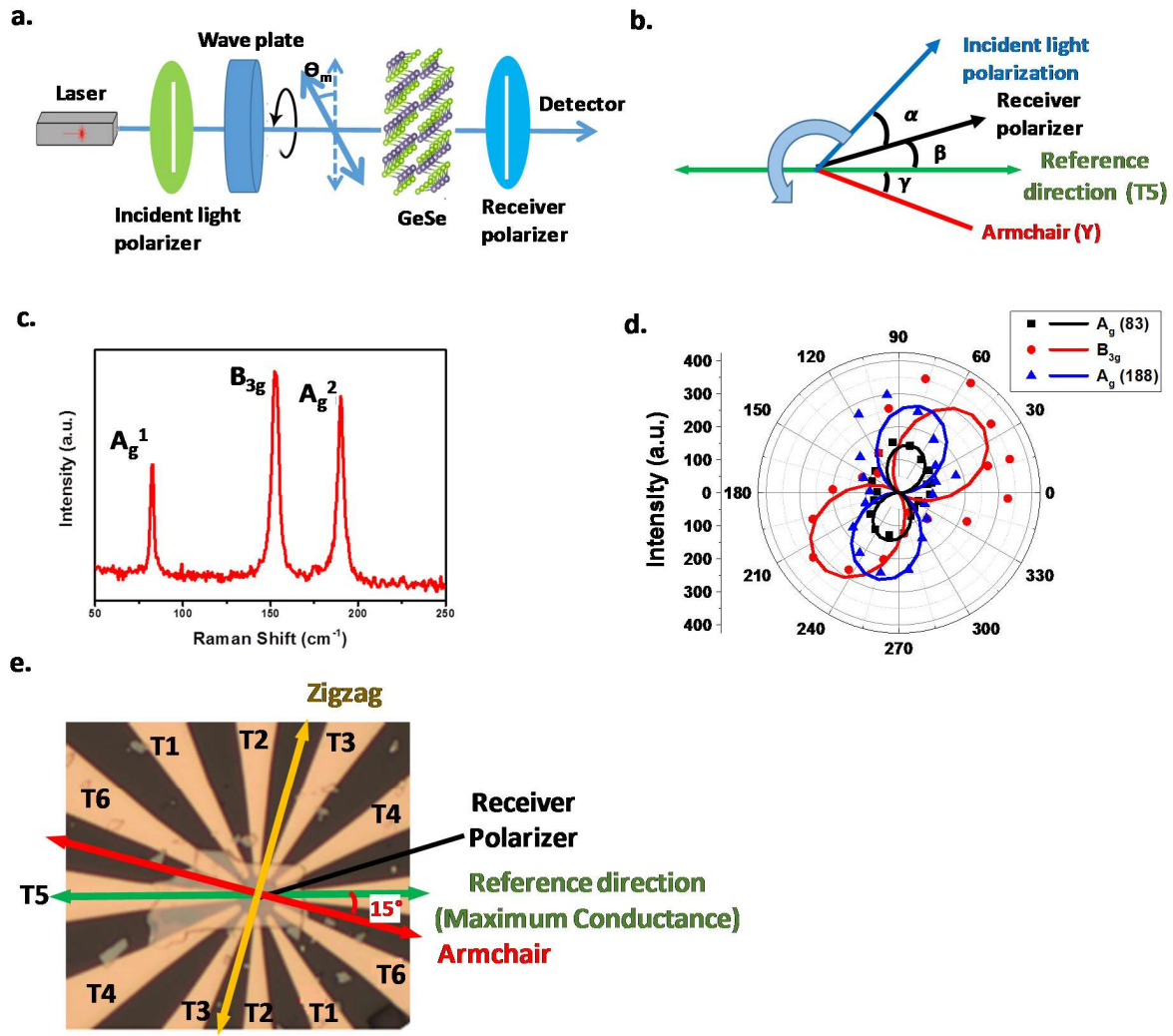

Figure 3. Angle resolved polarized Raman in GeSe. (a) Optical setup of the polarized Raman measurement. The half-wave plate can be rotated to modulate the angle between the polarization of the incident light and the reference direction, $\alpha$. (b) Illustration of the crystallographic axes and the direction of the polarization of the incident and scattered light, indicating the angle between the polarization of receiver polarizer and the armchair direction (Y axis), φ, and the angle between the polarization of the incident light and receiver polarizer, $\theta_m$. (c) A typical Raman spectrum of the GeSe. (d) Polar plots of the Raman intensities of $A_g$ and $B_{3g}$ modes as a function of the angle, $\alpha$. The symbols are measured results and the lines are calculations. (e) Optical image of the GeSe device. The armchair and zigzag direction determined by the polarized Raman are marked with the red and yellow lines respectively. The conductance reference direction along T5 (i.e. the maximum conductance direction) is marked with a green line.

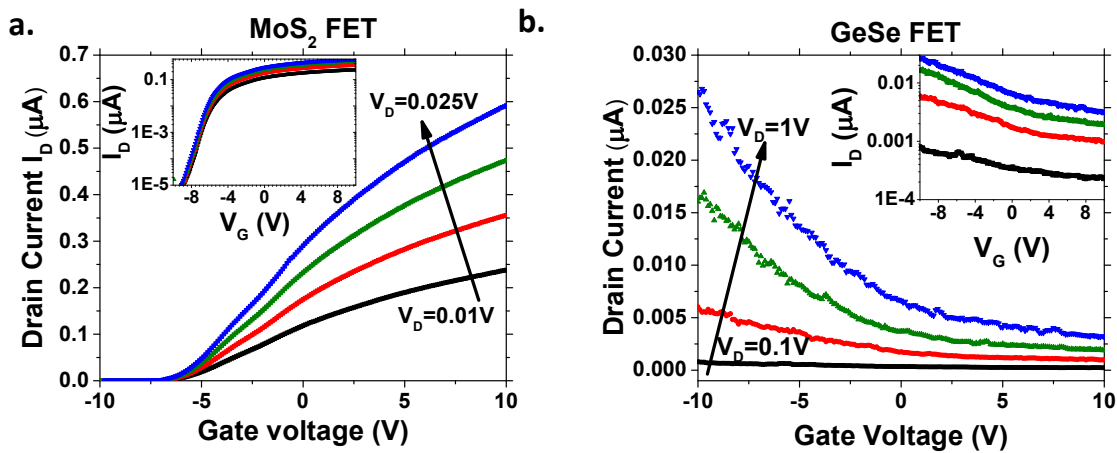

Figure 4. Transfer characteristics of (a) MoS$_2$ FET and (b) GeSe FET. I$_D$V$_G$ of MoS$_2$ transistor shows a typical n-channel behavior, while I$_D$V$_G$ of GeSe transistor shows a p-channel behavior. The semi-log scale plots of I$_D$V$_G$ characteristics are shown in the insets of each figures.

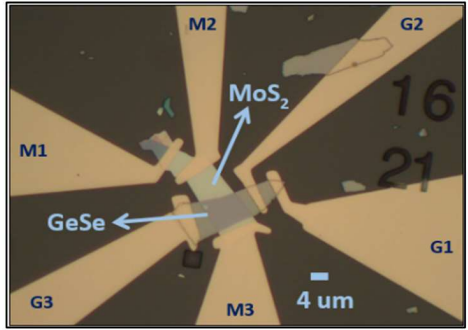
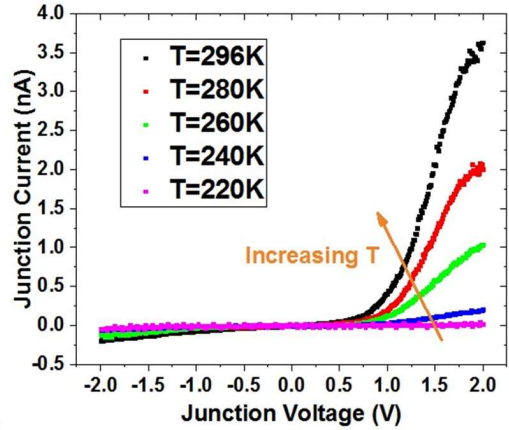
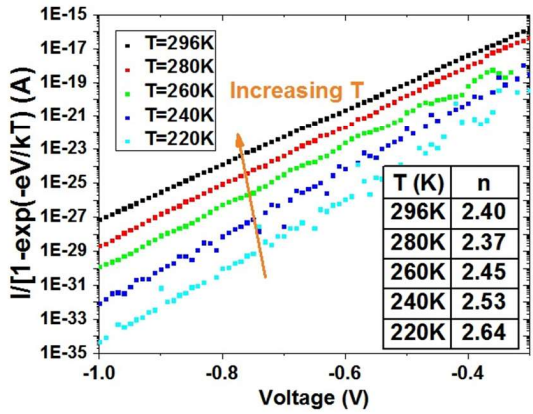
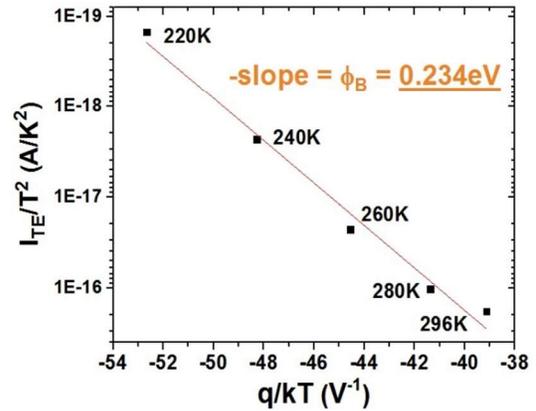

Figure 5 GeSe/MoS$_2$ heterostructure: (a) Optical image of a MoS$_2$ FET, GeSe FET, and MoS$_2$/GeSe junction device. The metal contacts to MoS$_2$ are labelled M1 to M3, and the metal contacts to GeSe are labelled G1 to G3. (b) I-V characteristics of the p-n heterojunction at different temperatures. (c) Extraction of ideality factor n from I-V curves of all temperature values. Slope gives $\frac{kT}{n}$ and intercept gives $\ln(I_{TE})$. (d) Extraction of band-offset $\phi_B = 0.234$eV from slope of semilog plot of $I_{TE}/T^2$ vs $q/kT$.